\documentclass{mn2e}

\usepackage{times}
\usepackage{graphicx}
\begin{document}

\title[Reionization history forecasts for Planck]
{Planck and reionization history: a model selection view}
\author[Pia Mukherjee and Andrew R. Liddle]
{Pia Mukherjee and Andrew R. Liddle\\ 
Astronomy Centre, University of Sussex, Brighton BN1 9QH, United
Kingdom}
\maketitle
\begin{abstract}
We use Bayesian model selection tools to forecast the Planck
satellite's ability to distinguish between different models for the
reionization history of the Universe, using the large angular scale
signal in the cosmic microwave background polarization spectrum. We
find that Planck is not expected to be able to distinguish between an
instantaneous reionization model and a two-parameter smooth
reionization model, except for extreme values of the additional
reionization parameter. If it cannot, then it will be unable to
distinguish between different two-parameter models either. However,
Bayesian model averaging will be needed to obtain unbiased estimates
of the optical depth to reionization. We also generalize our results
to a hypothetical future cosmic variance limited microwave anisotropy
survey, where the outlook is more optimistic.
\end{abstract}

\section{Introduction}

The five-year data from the Wilkinson Microwave Anisotropy Probe
(WMAP: Hinshaw et al.~2008; Dunkley et al.~2008; Komatsu et al.~2008)
have given reasonably tight constraints on the optical depth to
Thomson scattering from the last-scattering surface, $\tau = 0.09 \pm
0.02$ with modest dependence on inclusion of additional datasets and
changes to model assumptions. It has not however had the accuracy
needed to go beyond this one-parameter description of the ionization
history of the Universe, to give a more detailed view of how
reionization took place and to distinguish between the various models
in the literature (though combined with tentative indication of a
change in Lyman-$\alpha$ optical depth around redshift 7, it does give
some indication that reionization is an extended process).

Theoretical studies suggest that the process of reionization can be
quite complex (eg Barkana \& Loeb 2001; Haiman \& Holder 2003; Cen
2003; for reviews see Barkana \& Loeb 2007a; Meiksin 2007).  The Planck
satellite may have the sensitivity to go beyond a one-parameter
description of the process.  For instance, Lewis, Weller \& Battye
(2006) considered three specific reionization histories (with other
cosmological parameters held fixed), and assessed whether Planck would
be able to distinguish amongst them, finding that it did indeed have
some ability to do so.

However, the true data analysis problem is more complicated than in
their study. Future experiments will not be trying to distinguish
between a small set of specific reionization histories. Rather, there
will be competing \emph{models} for reionization each of which feature
parameters that need to be determined from the data. That is, the
problem is one of \emph{model selection} (see Gregory 2005; Liddle,
Mukherjee \& Parkinson 2006a; Trotta 2008, and references therein).
In this paper we use Bayesian model selection tools to forecast the
ability of the Planck satellite, and a putative
cosmic-variance-limited future survey, to distinguish between two
reionization models, instantaneous reionization (parameterized solely
by the optical depth to reionization $\tau$ or equivalently the
redshift of reionization), and a smooth transition to the ionized
state, parameterized by a further parameter $d_\eta$ which measures the
rapidity of the transition (in conformal time $\eta$).

We consider only the large-scale bump in the cosmic microwave
background (CMB) polarization spectrum generated by Thomson scattering
of the CMB quadrupolar anisotropy during reionization. The detailed
shape of the bump is related to the evolution of the globally-averaged
ionized fraction during reionization (Kaplinghat et al.~2003; Hu \&
Holder 2003; Colombo et al.~2005).  The power on scales smaller than
the horizon size at reionization is uniformly damped by $e^{-2\tau}$;
this then cannot be used to constrain the details of reionization
beyond $\tau$, or even to constrain $\tau$ itself which would be
almost completely degenerate with the amplitude of
perturbations. Other degeneracies are discussed in Martins et
al.~(2004) and Trotta \& Hansen (2004).  Reionization affects the CMB
spectrum again on much smaller scales via secondary effects due to
inhomogeneous or patchy reionization and the Ostriker--Vishniac
effects (Ostriker \& Vishniac 1986; Weller 1999; Hu 2000). We do not
consider these effects which are beyond multipole $\ell\sim2000$,
modelling only uniform reionization. In the future, 21cm emission from
neutral hydrogen is expected to provide a good tracer of the details
of reionization (eg. see Barkana \& Loeb 2007b), and there are
experiments that will focus on mapping this emission.

\begin{figure*}
\begin{tabular}{cc}
\includegraphics[width=7cm]{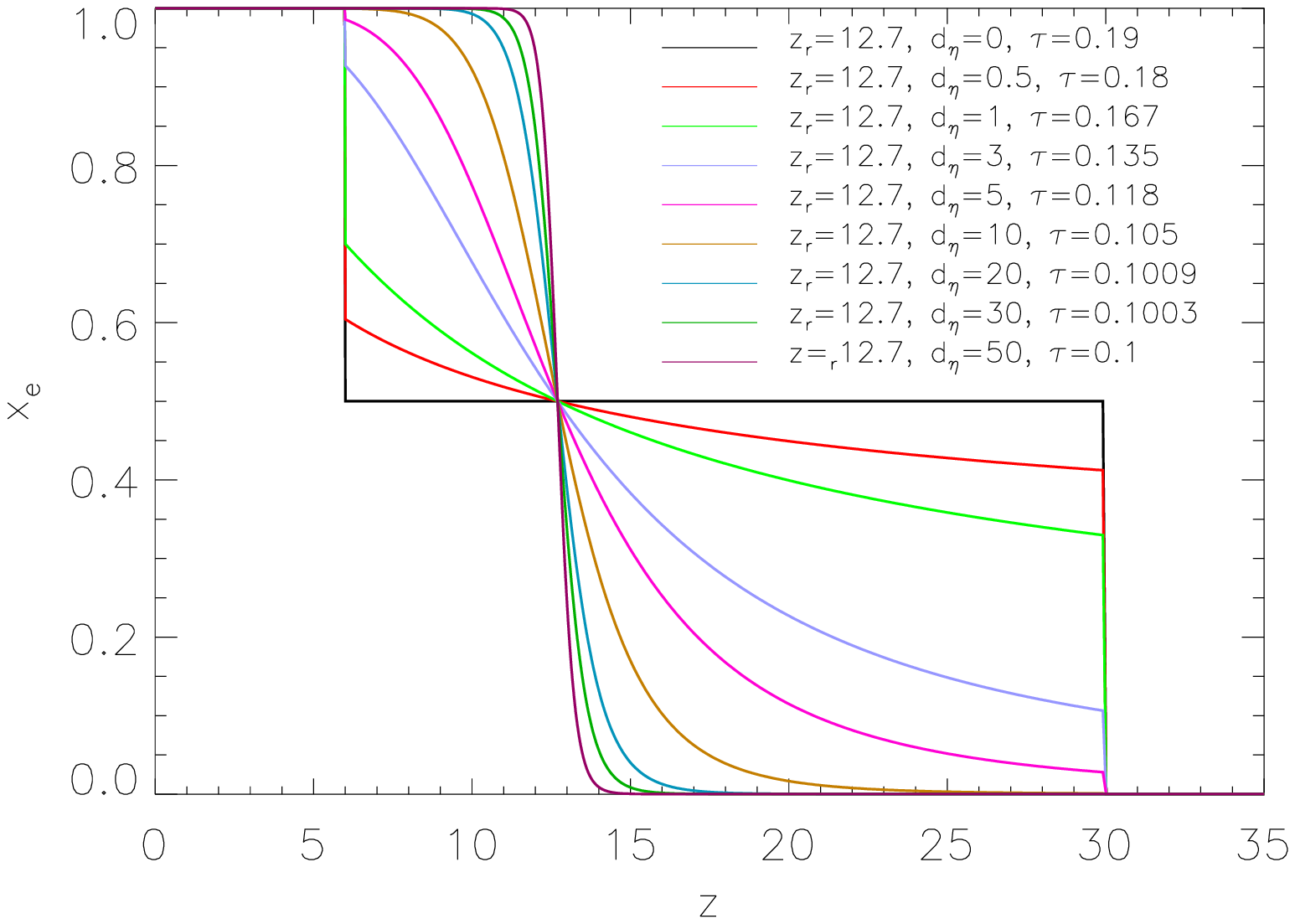} &
\includegraphics[width=7cm]{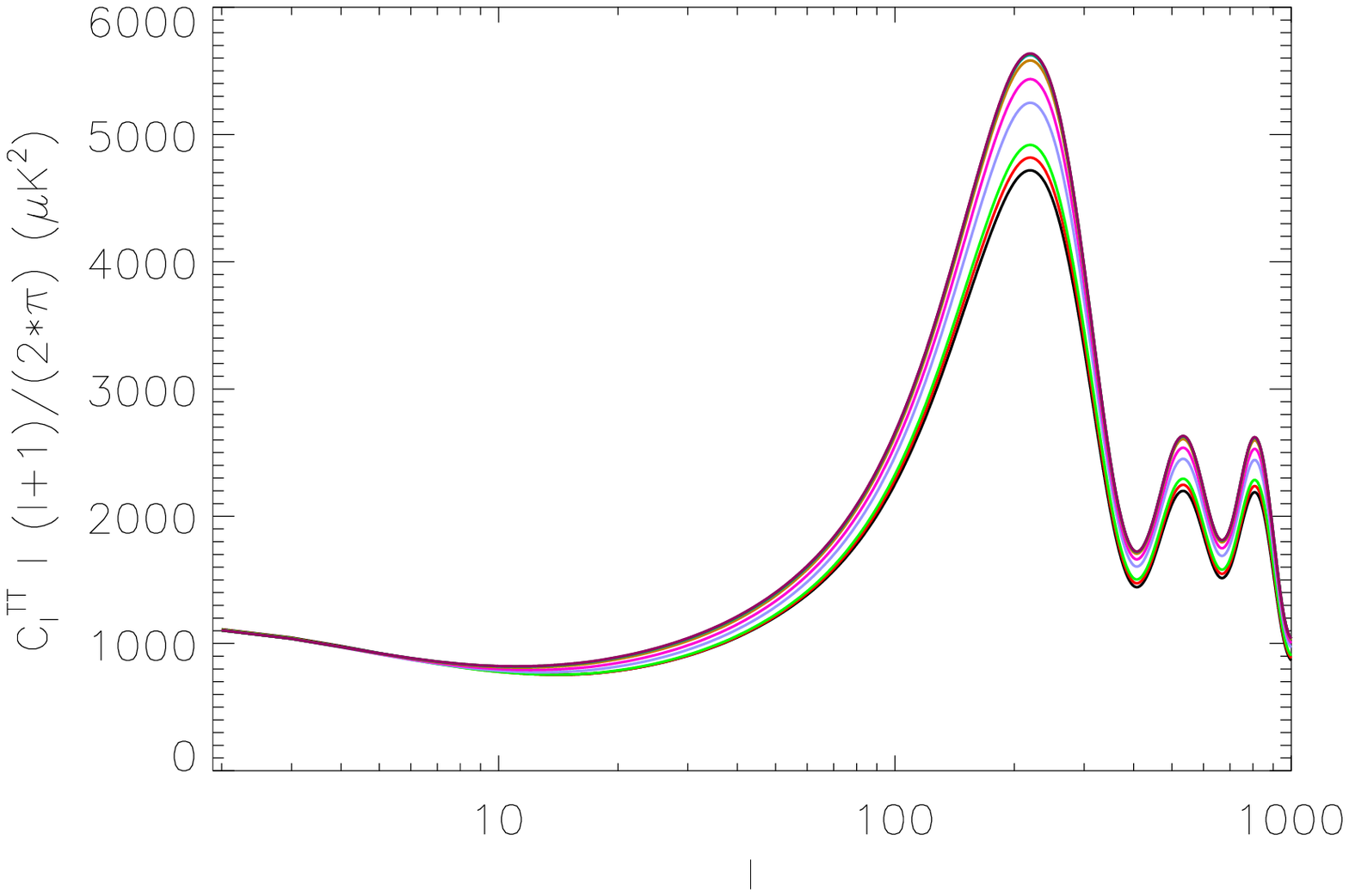}\\
\includegraphics[width=7cm]{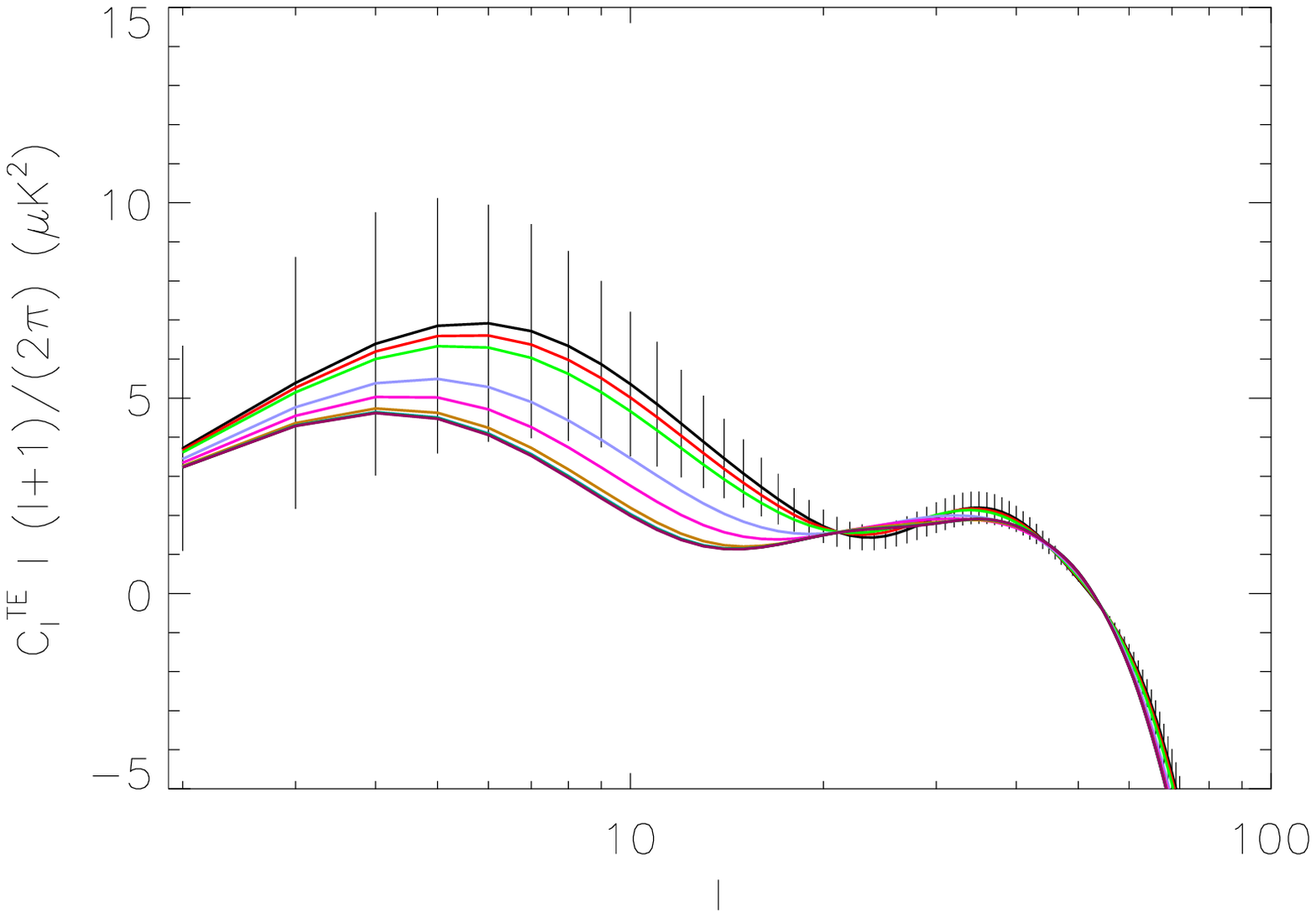} &
\includegraphics[width=7cm]{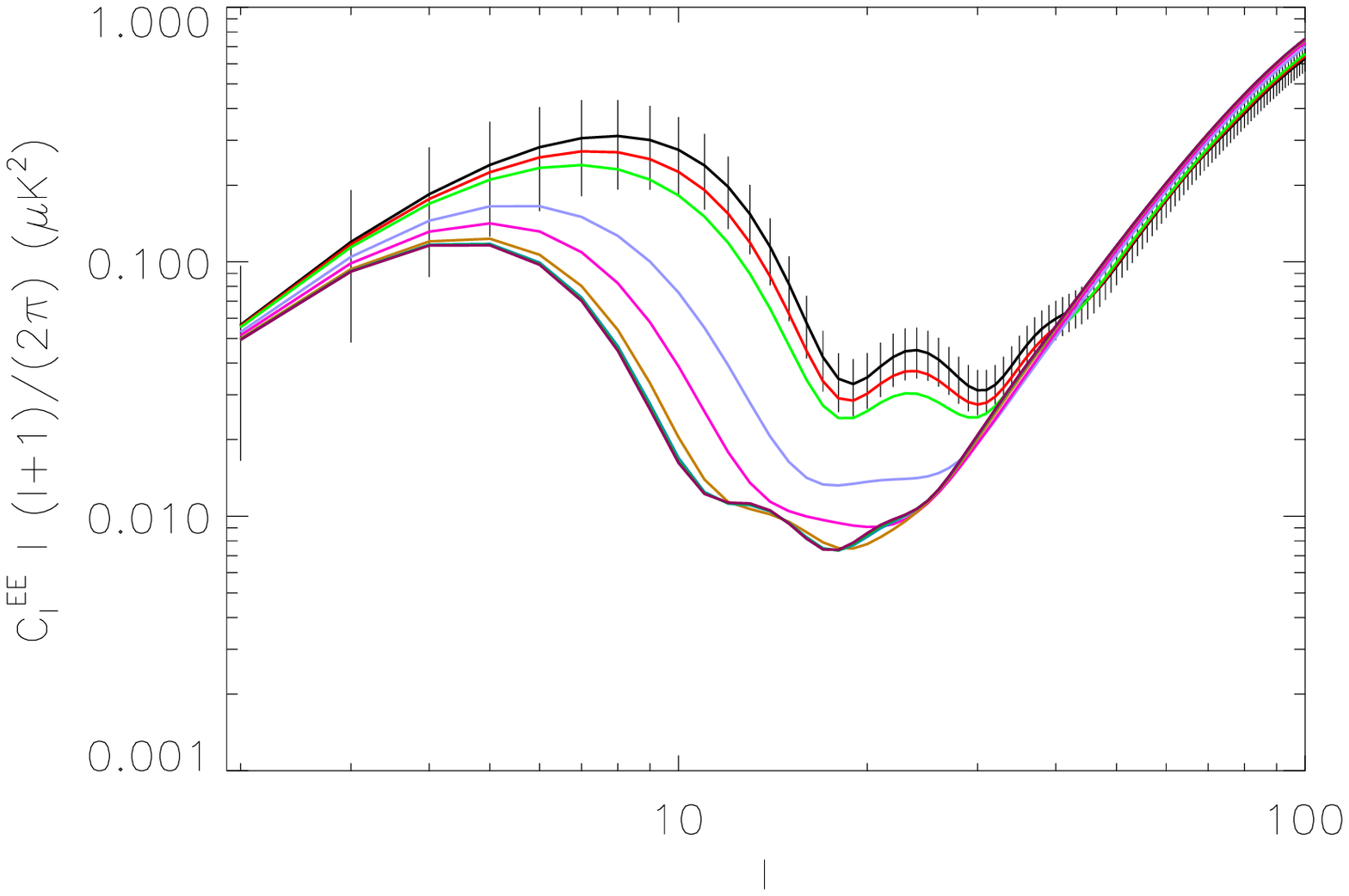}\\
\end{tabular}
\caption{\label{f:models3} The upper left panel shows a series of
  reionization histories, in each case with the reionization redshift
  fixed to $z_{\rm r} = 12.7$. The remaining panels show the TT (top right),
  TE (bottom left) and EE (bottom right) spectra generated by these
  models. The error bars indicate the uncertainty due to cosmic
  variance.}
\end{figure*}

\section{The models}

Our cosmological model is the usual spatially-flat $\Lambda$CDM
cosmology, seeded by power-law adiabatic density perturbations. Its
adjustable parameters are the dark matter and baryon densities
$\Omega_{\rm c}$ and $\Omega_{\rm b}$, the Hubble parameter $h$, and
the perturbation amplitude $A_{\rm s}$ and spectral index $n_{\rm
s}$. These are fixed to WMAP3 best-fit values\footnote{Our
calculations predated the WMAP five-year data release (Hinshaw et
al.~2008), which however left the numbers almost unchanged.} (Spergel
et al.~2007) for $\Omega_{\rm b} h^2$, $\Omega_{\rm c} h^2$, the
projected sound horizon $\theta$, $A_s \exp(-2\tau)$ and $n_s$. We
then study the reionization signal from the TE and EE spectra out to
$\ell$ of 100.  It is possible to use such an analysis procedure
because the non-reionization parameters are very well determined by
the TT spectrum, and because the large-scale signal in CMB
polarization is independent of the other parameters. A similar
procedure has been followed in works including Kaplinghat et
al.~(2003), Holder et al.~(2003), and Mortonson \& Hu (2008a,b). The
uncertainty on $\tau$ derived holding these parameters fixed is
expected to be an underestimate by about 10\% (Mortonson \& Hu 2008a).

We assume standard recombination. If the recombination model
eventually needs to be modified to account for two-photon decays
(Dubrovich \& Grachev 2005; Wong \& Scott 2007; Chluba \&
Sunyaev 2008; Hirata 2008), this should not affect the model
comparisons we present here because it would be common to all the
models. In addition, the spectrum changes on intermediate to small
scales while we are using only the large scales here.

We mainly consider a two-parameter reionization model defined by the
ionization fraction history 
\begin{equation}
x_e(\eta_i) = \left[1-x_e(\eta_{i-1})\right]\,\frac{\tanh\left[
\left(\frac{\eta}{\eta_{z_{\rm r}}}-1\right)d_{\eta}+1\right]}{2} +
x_e(\eta_{i-1}).
\end{equation}
where $x_e$ refers to the ionization fraction, $\eta_i$ and
$\eta_{i-1}$ refer to consecutive time steps, $\eta$ to the conformal
time at the $i$-th time step, $z_{\rm r}$ is the redshift at which the
ionization fraction is 0.5, $\eta_{z_{\rm r}}$ is the conformal time
corresponding to that redshift, and $d_{\eta}$ gives the (inverse)
width of the transition. Such a transition is implemented in CAMB
(Lewis, Challinor \& Lasenby 2000), and the commonly-used
instantaneous reionization scenario corresponds to $d_{\eta}$ having a
large enough value, such as 50, that $z_{\rm r}$ is effectively the
redshift of instantaneous reionization.\footnote{We use a version of
CAMB prior to April 2008, that includes only hydrogen reionization and
thus a final ionization fraction of unity. Including helium
reionization the final ionization fraction would be $\sim1.08$. See
section XIII of CAMB notes, via a link from
http://camb.info/readme.html, for further details. The model selection
results presented in this paper are not expected to change following
this inclusion of helium reionization.}

\begin{figure*}
\begin{tabular}{cc}
\includegraphics[width=7cm]{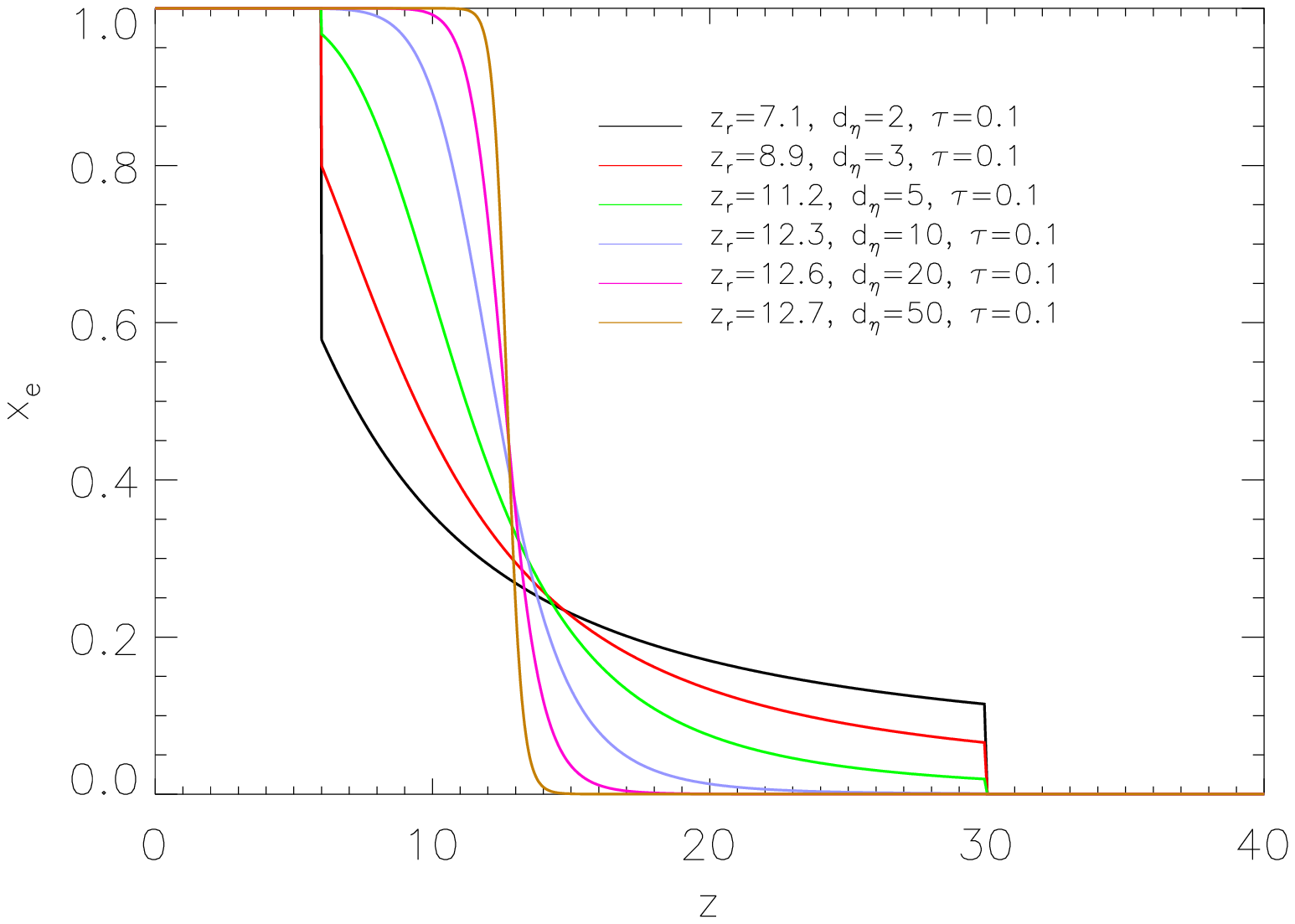} &
\includegraphics[width=7cm]{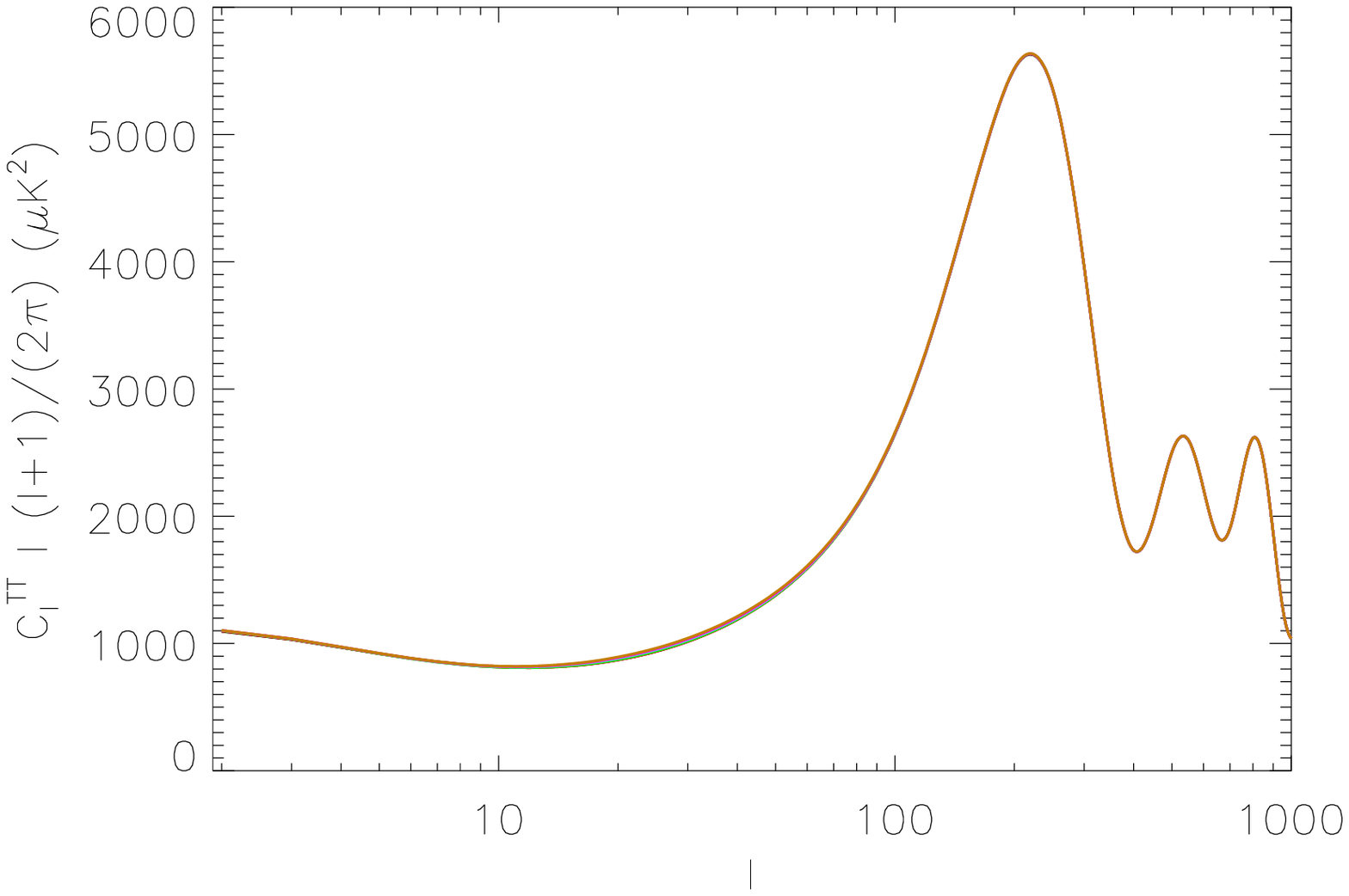}\\
\includegraphics[width=7cm]{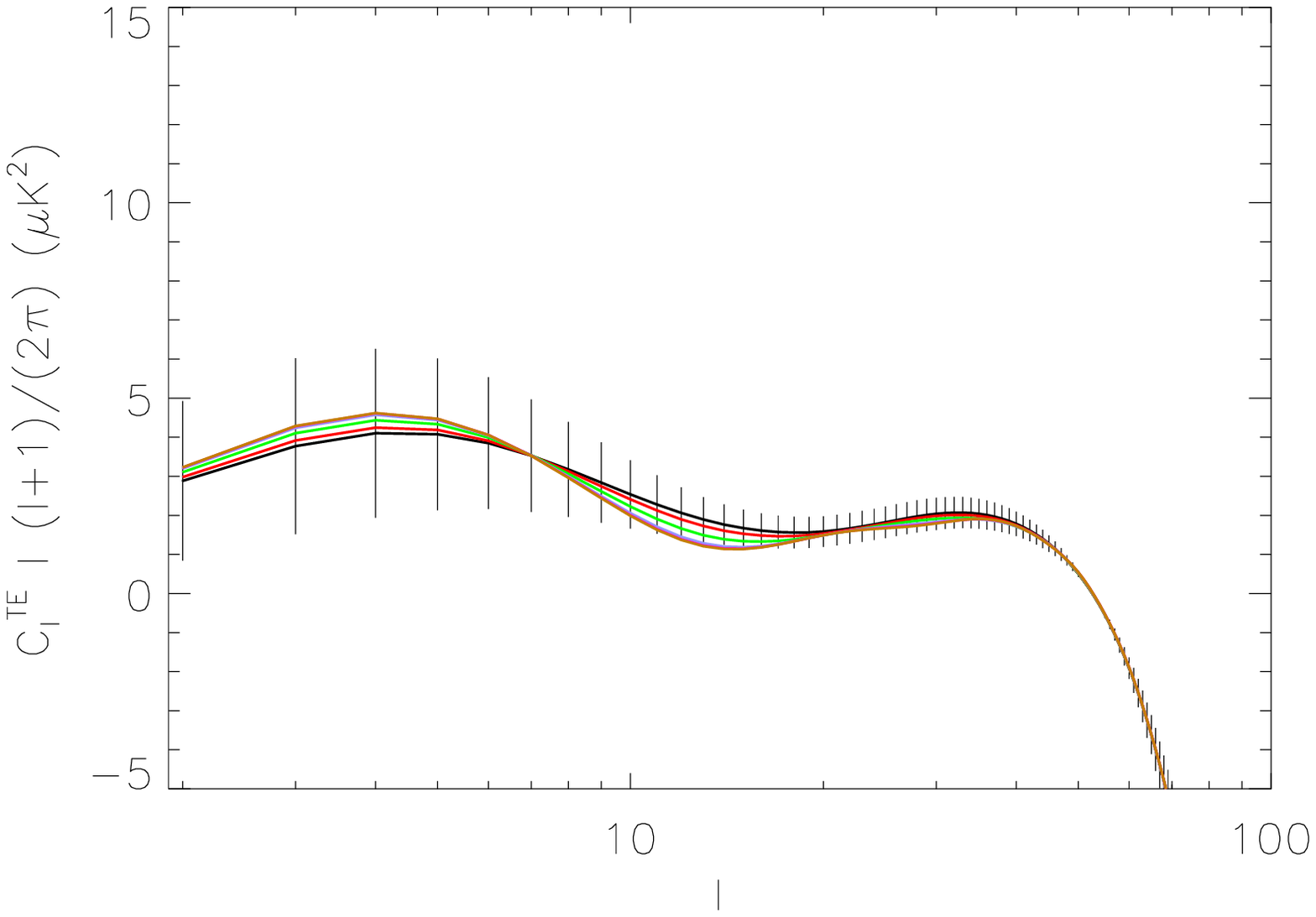} &
\includegraphics[width=7cm]{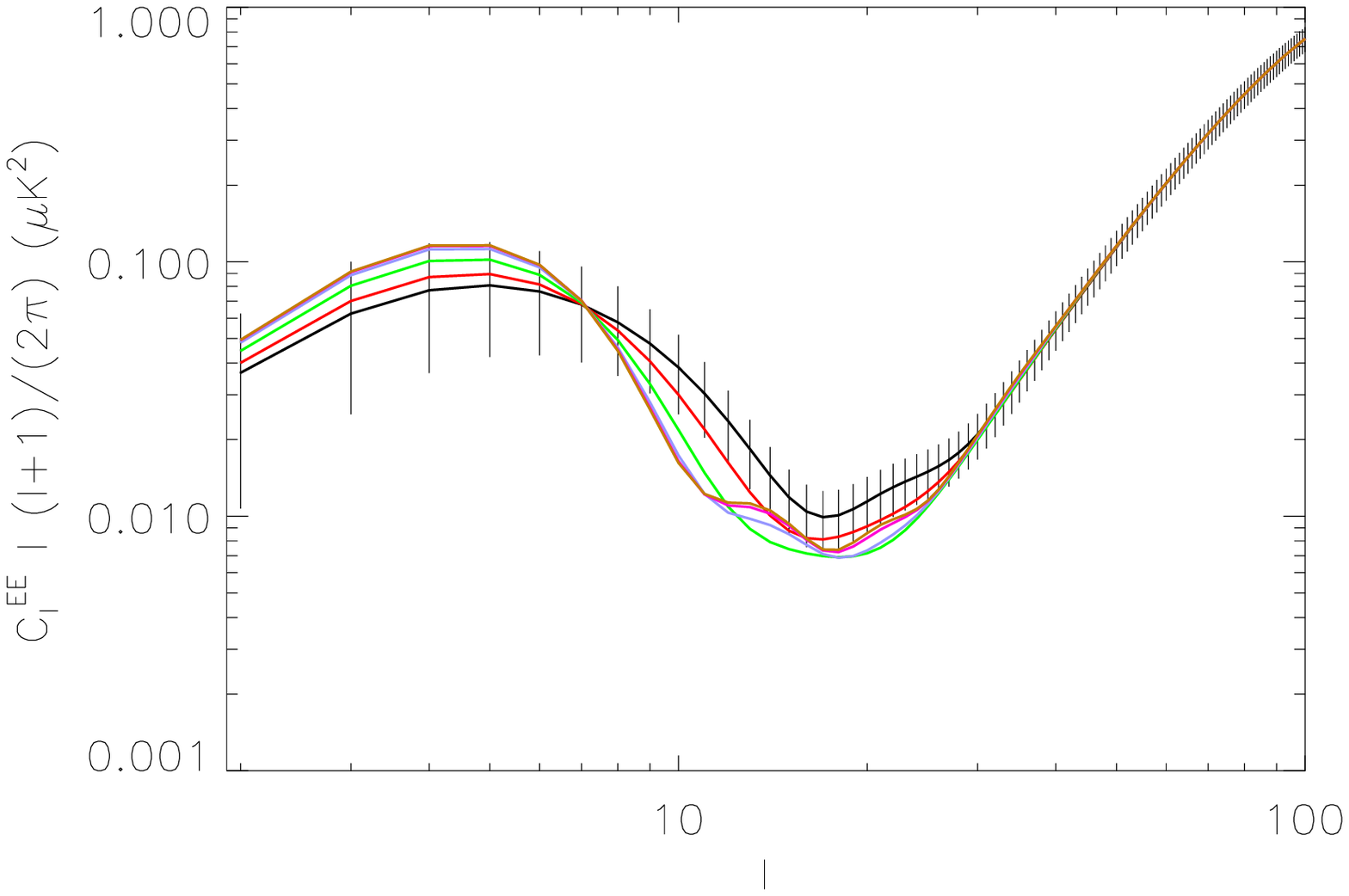}\\
\end{tabular}
\caption{\label{f:models4} As Figure~\ref{f:models3}, but for models
each with $\tau$ fixed at 0.1. At fixed $\tau$ the spectra, especially
TT, have only a weak dependence on $d_\eta$.}
\end{figure*}

We additionally force the ionization fraction to unity for
\mbox{$z<6$}, to avoid conflict with quasar absorption spectrum data,
and to zero for $z>30$ as no ionizing sources are expected so early.
The optical depth $\tau$ is computed numerically for any such
reionization history.  Given the reionization history, we compute the
CMB power spectra using a version of CAMB with minor
modifications. Our assumed cosmological model has only scalar initial
perturbations, and so we do not compute the BB polarization spectra.

Figures \ref{f:models3} and \ref{f:models4} show some predicted power
spectra for these models, showing in particular that $\tau$ is indeed
mainly responsible for variations in the predictions and hence the
most readily measured parameter. Furthermore, at fixed $\tau$ it is
clear from figure \ref{f:models4} that all the discriminating power is
in the polarization spectra rather than temperature.

Through most of the following analysis we take a fiducial \mbox{$z_{\rm
r}$--$d_\eta$} model with $d_\eta=3$ and $z_{\rm r}=8.9$, corresponding
to $\tau=0.1$ as already well determined by WMAP5.

\begin{figure}
\centering
\includegraphics[width = 7 cm]{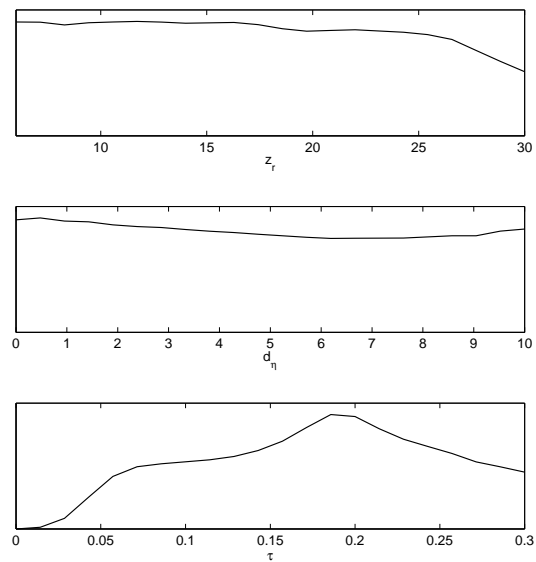}
\caption{\label{f:testpriors} Uniform priors on $z_{\rm r}$ between 6
and 30, and on $d_\eta$ between 0 and 10, result in a non-uniform
prior on $\tau$ (solid curves). We can work with such a prior on
$\tau$ because expected uncertainties from a Planck-like experiment
are $\Delta \tau=0.01$ and over such a range the prior is fairly
flat.}
\end{figure}

Flat priors are assumed on $z_{\rm r}$ and $d_\eta$, over ranges of
6--30 and 0--10 respectively (the figures show that values of $d_\eta$
larger than this result in scenarios very close to
instantaneous). Figure \ref{f:models3} also motivates us to try a
logarithmic prior on $d_\eta$, for which we take the range
0.3--30. However a $\tau<0.3$ prior is also imposed, so that the one
additional parameter as compared to instant reionization does not
correspond to one additional degree of freedom. For this reason
regular likelihood ratio tests, of the kind performed in Kaplinghat et
al.~(2003) and Holder et al.~(2003), will not be valid. Figure
\ref{f:testpriors} shows the induced prior on $\tau$ resulting from
linear priors on $z_{\rm r}$ and $d_\eta$ (their own priors are not
perfectly flat due to the extra imposition of the $\tau < 0.3$
prior). The uncertainty on $\tau$ from Planck is of order 0.01, and
over such widths the prior on $\tau$ is roughly uniform; it is even
more so around the fiducial value of $\tau$. The same applies to the
case of a log prior on $d_\eta$.

\begin{table*}
\centering
\begin{tabular}{l|c|c|c|c|}
\hline
& Model, priors & parameter estimates & ln Evidence & $\Delta$ ln E \\ 
\hline \hline 
& instantaneous reionization & $z_{\rm r}=12.9\pm 0.5$ &
$-6.3\pm0.1$ & 0.0\\  
& $z_{\rm r}$: 6--30, $d_\eta=50$  & $\tau=0.108\pm0.006$ & & \\ \hline
Planck satellite & linear $d_\eta$ model & $z_{\rm r}=10.1\pm1.7$,
$d_\eta=4.4\pm1.9$ & 
$-4.4\pm0.2$ & 1.9 \\
& $z_{\rm r}$: 6--30, $d_\eta$: 0--10 & $\tau=0.103\pm0.006$ & & \\ \hline
& log $d_\eta$ model & $z_{\rm r}=9.9\pm1.9$, $d_\eta=4.1\pm1.9$ &
$-4.7\pm0.2$ & 1.6 \\  
& $z_{\rm r}$: 6--30, log $d_\eta$: -1.2--3.4 & $\tau=0.103\pm0.006$ & & \\ 
\hline \hline 
& instantaneous reionization & $z_{\rm r}=12.7\pm 0.2$ &
$-15.8\pm0.1$ & 0.0 \\  
& $z_{\rm r}$: 6--30, $d_\eta=50$  & $\tau=0.106\pm0.003$ & & \\ \hline
cosmic variance & linear $d_\eta$ model & $z_{\rm r}=9.4\pm1.0$,
$d_\eta=3.3\pm0.5$ & 
$-6.6\pm0.1$ & 9.2 \\
limited & $z_{\rm r}$: 6--30, $d_\eta$: 0--10 & $\tau=0.102\pm0.005$ &
& \\ \hline 
& log $d_\eta$ model & $z_{\rm r}=9.2\pm1.0$, $d_\eta=3.2\pm0.4$ &
$-6.8\pm0.3$ & 9.0 \\  
& $z_{\rm r}$: 6--30, log $d_\eta$: -1.2--3.4 & $\tau=0.101\pm0.005$ &
& \\  
\hline 
\end{tabular}
\caption{Analyzing TE and EE spectra of Planck specifications (first
panel), with a fiducial model of $z_{\rm r}=8.9$ and $d_\eta=3$
(implying $\tau=0.1$) using three test models. The second panel shows
the same for a cosmic variance limited experiment.  ln Evidences are
based on four estimates of the evidence for each model.}
\label{table1}
\end{table*}

Other models of reionization have been proposed (Haiman \& Holder
2003; Cen 2003), for instance the double reionization scenario
considered in Lewis et al.~(2006), which would in general require a
third parameter. From a model selection point of view (i.e.\ taking
into account parameter uncertainties within each model) results in
this paper indicate that discriminating such a model from a smooth
transition model would be beyond the scope of Planck, though perhaps
within the scope of a closer to cosmic variance limited
experiment. Here our focus is mainly on clarifying what Planck can
learn about reionization.

\section{Model selection forecasting methodology}

Model selection forecasting assesses a given experiment's ability to
distinguish between different cosmological models. This ability
necessarily depends on the true model and on its parameter values (and
of course on the overriding assumption that one of the models we will
be considering is, if not the actual true model, at least
representative of its predictions for the experiment under
consideration). We call this true model and its parameter values the
\emph{fiducial model}.  Trotta (2007) introduced an approach to model
selection forecasting, PPOD, which averaged the model selection
forecast over the current knowledge of model parameters, so as to give
the probability of the future experiment giving different model
selection outcomes. Mukherjee et al.~(2006b) adopted a different
approach where the model selection outcome was forecasted as a
function of the fiducial model parameters, so as to assess where in
the parameter space the experiment could strongly distinguish the
models, and defined some experimental figures of merit based on this
notion.

Computational restrictions prevent us from making an extensive
investigation of how the model selection outcome will depend on the
fiducial model chosen. Instead, we consider a single fiducial $z_{\rm
r}$--$d_\eta$ model with $d_\eta=3$ and $z_{\rm r}=8.9$, corresponding
to $\tau=0.1$, and simulate Planck quality data for it. Figure
\ref{f:models4} shows that this model is quite different from an
instantaneous reionization model with the same $\tau$ (though models
can be constructed that are even more drastically different, up to the
step model seen in Figure 1).  Our goal is to determine whether Planck
could distinguish the two-parameter model from an instantaneous
reionization model, for these fiducial parameters with a certain level 
of strength of evidence. If it can, then the
threshold for detection lies between this model and the instantaneous
reionization model, otherwise it lies further away.

Throughout we use the nested sampling algorithm for computing model
evidences, first implemented for cosmological applications in
Mukherjee, Parkinson \& Liddle (2006a) and subsequently developed in
Parkinson, Mukherjee \& Liddle (2006). It was used to make forecasts
for Planck's ability to determine inflationary parameters in Pahud et
al.~(2006,2007). This algorithm, due to Skilling (2006), computes the
Bayesian evidence for any given model, as well as providing parameter
estimates within that model. The evidence is the probability of the
data given the model, hence can be used to determine how likely each
model is to have given rise to the data. A difference of 2.5 in log
evidence can be taken to be significant, and 5 decisive, evidence in
favour of the model with larger evidence (Jeffreys 1961).

\section{Results}

\subsection{The Planck satellite}

We model Planck TE and EE data using just the 143 GHz polarization
channel, following for its specifications the current Planck
documentation.\footnote{
www.rssd.esa.int/index.php?project=PLANCK\&page=perf\underline{~}top}
The full likelihood is constructed in the manner of Lewis (2005) and
Pahud et al.~(2006,2007), without creating noisy data
realizatons. This ensures that the bias issues we discuss below are
not a result of realization noise, but instead the forecast is
equivalent to averaging over many data realizations and is thus itself
effectively unbiased (see the appendix of Sahl\'en et al.~2008). We
assume a sky coverage of 0.8, and take the likelihood up to a maximum
multipole of $\ell = 100$.

\begin{figure*}
\begin{tabular}{ccc}
\includegraphics[width = 6 cm]{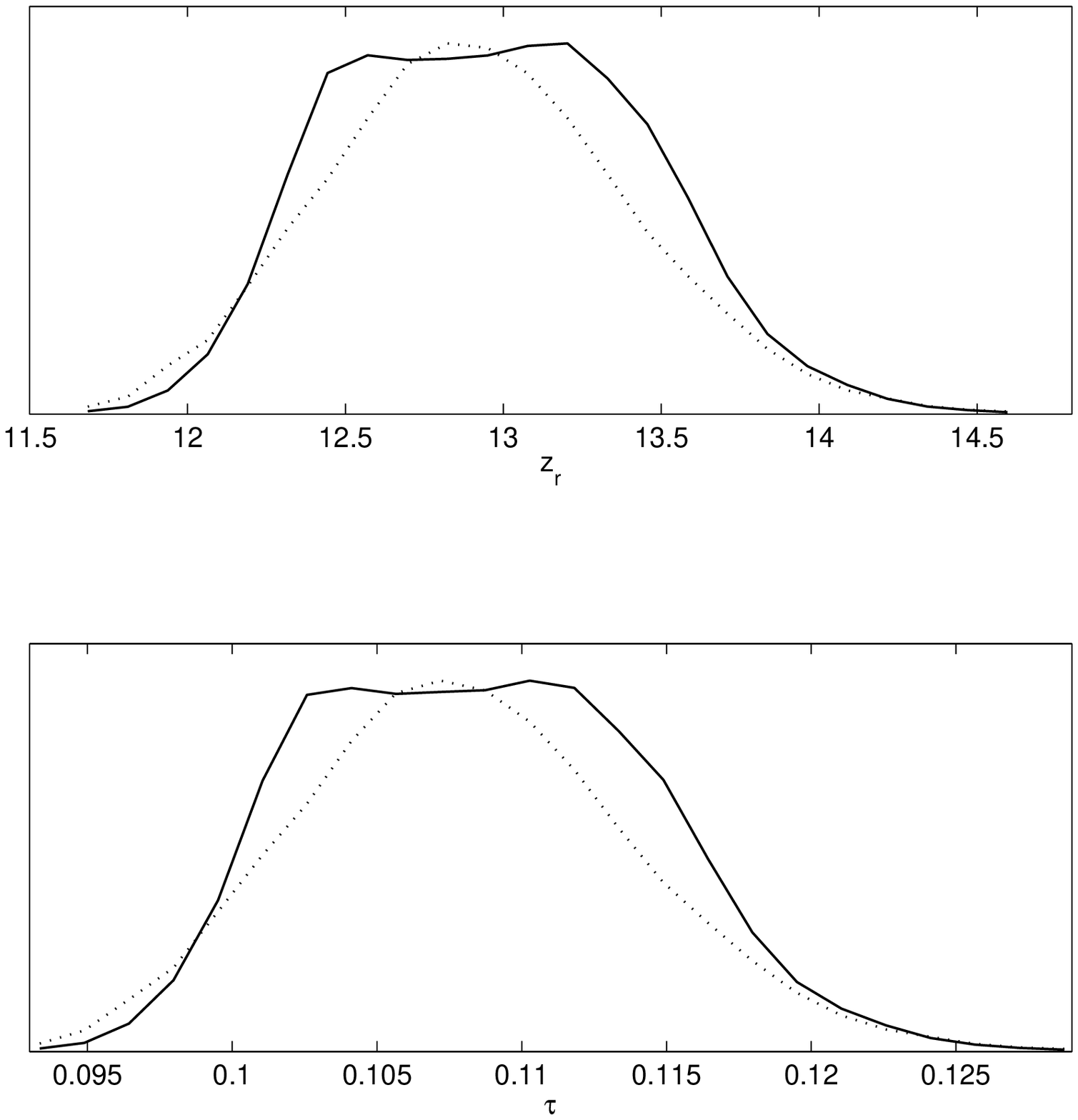}
\includegraphics[width
  =6cm]{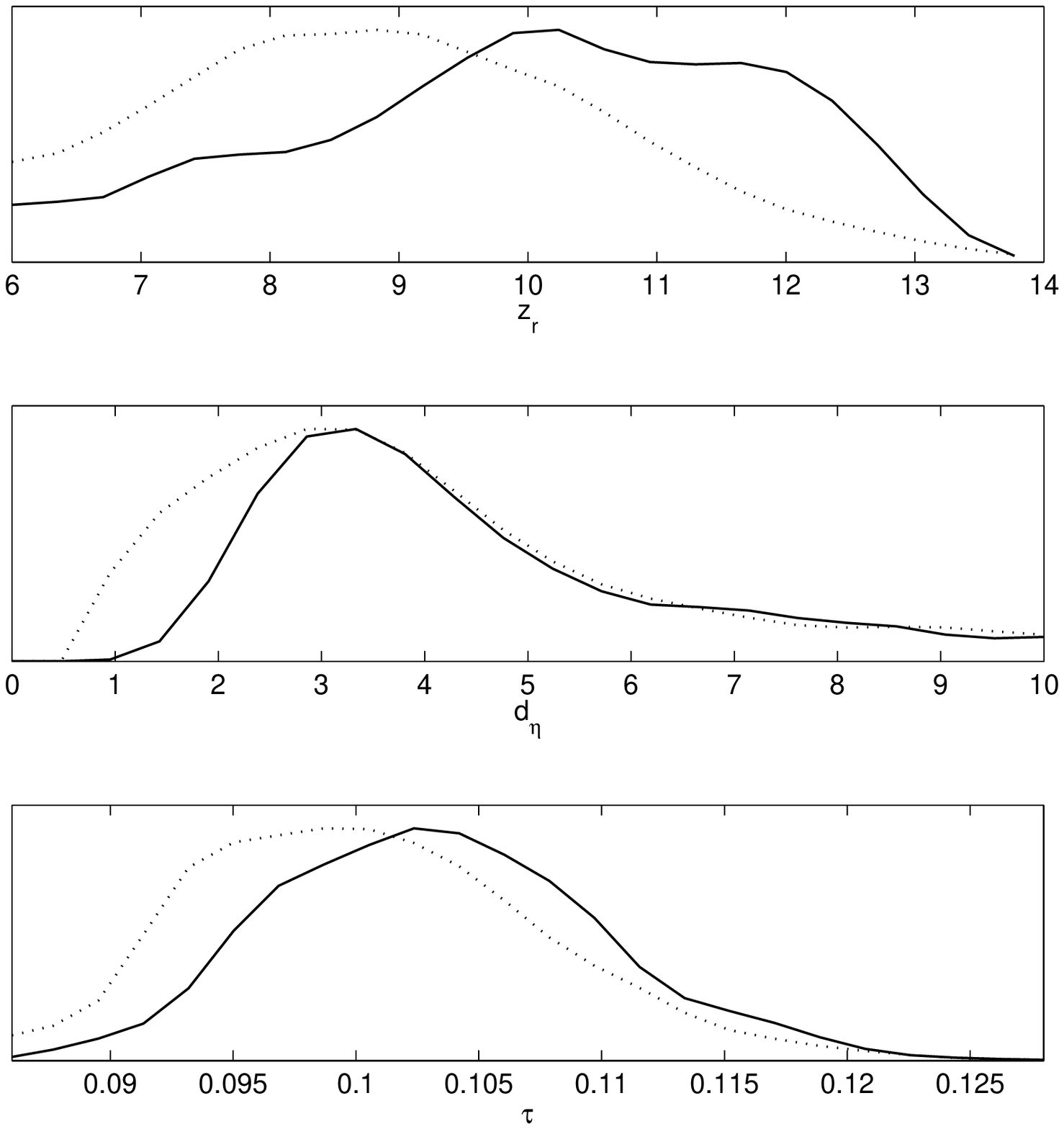} 
\includegraphics[width =
  6cm]{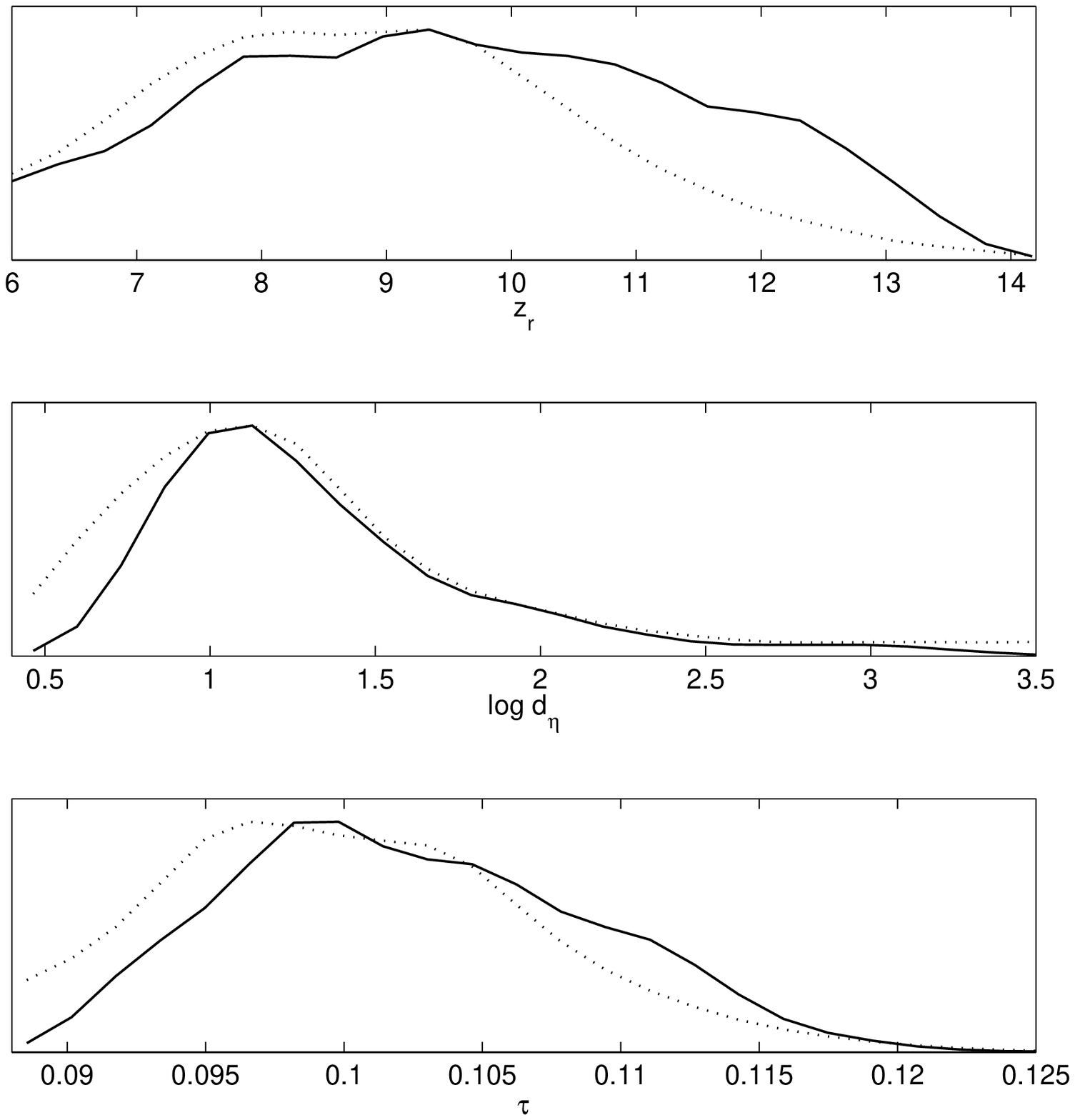} 
\end{tabular}
\caption{\label{f:paramspl} A fiducial model with $z_{\rm r}=8.9$ and
  $d_\eta=3$ treated with an instantaneous reionization model
  ($d_\eta$ fixed at 50, $z_{\rm r}$ allowed to vary between 6 and 30)
  (left panel), a $z_{\rm r}$--$d_\eta$ reionization model with
  $z_{\rm r}$ between 6 and 30 and $d_\eta$ between 0 and 10 (centre
  panel), and a $z_{\rm r}$--$d_\eta$ reionization model with log
  prior in $d_\eta$ ($z_{\rm r}$ between 6--30, log $d_\eta$ between
  -1.2 and 3.4 corresponding to 0.3 and 30) (right panel).}
\end{figure*}

\begin{figure*}
\begin{tabular}{ccc}
\includegraphics[width = 6 cm]{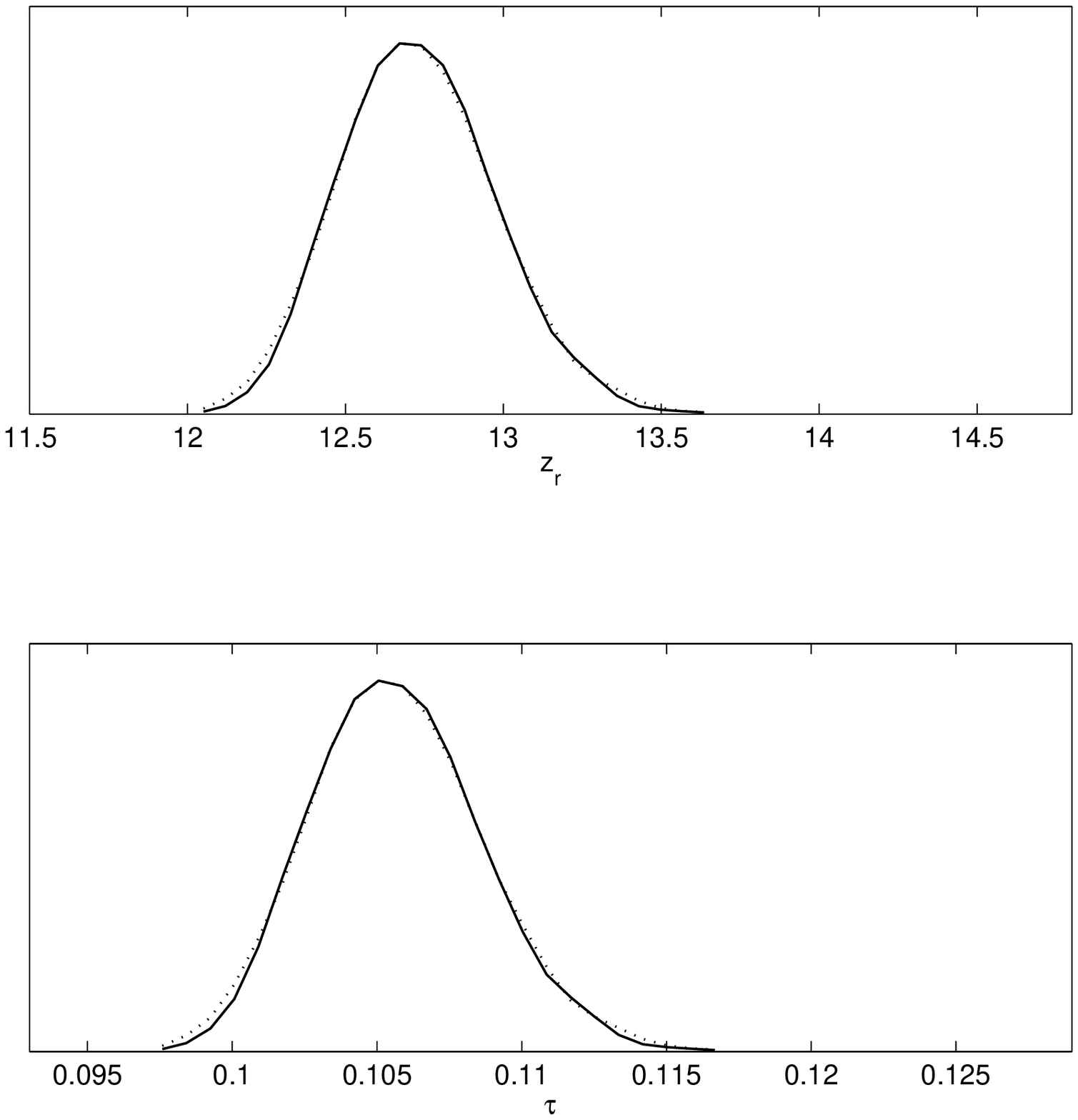}
\includegraphics[width = 6
  cm]{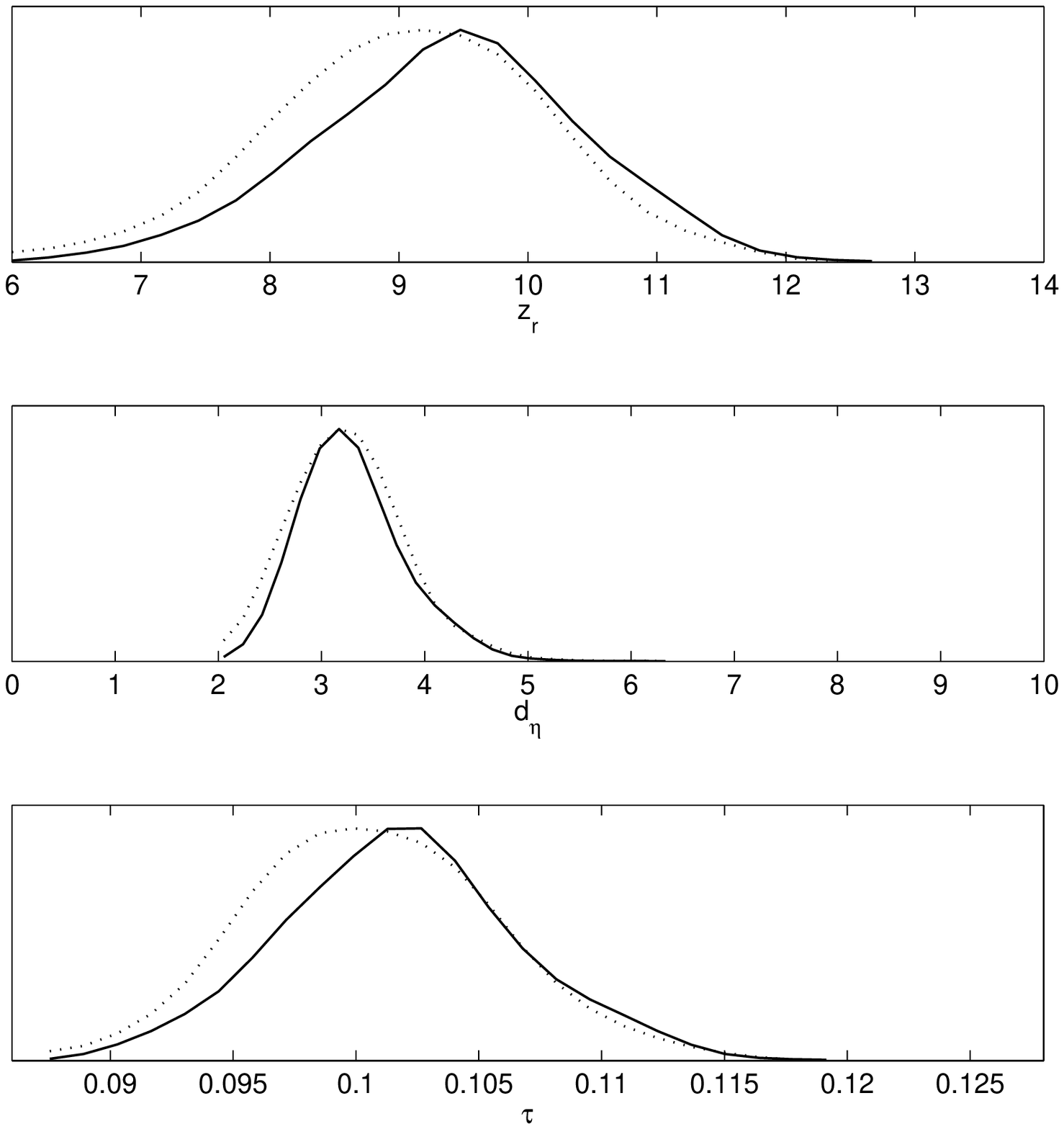}
\includegraphics[width = 6
  cm]{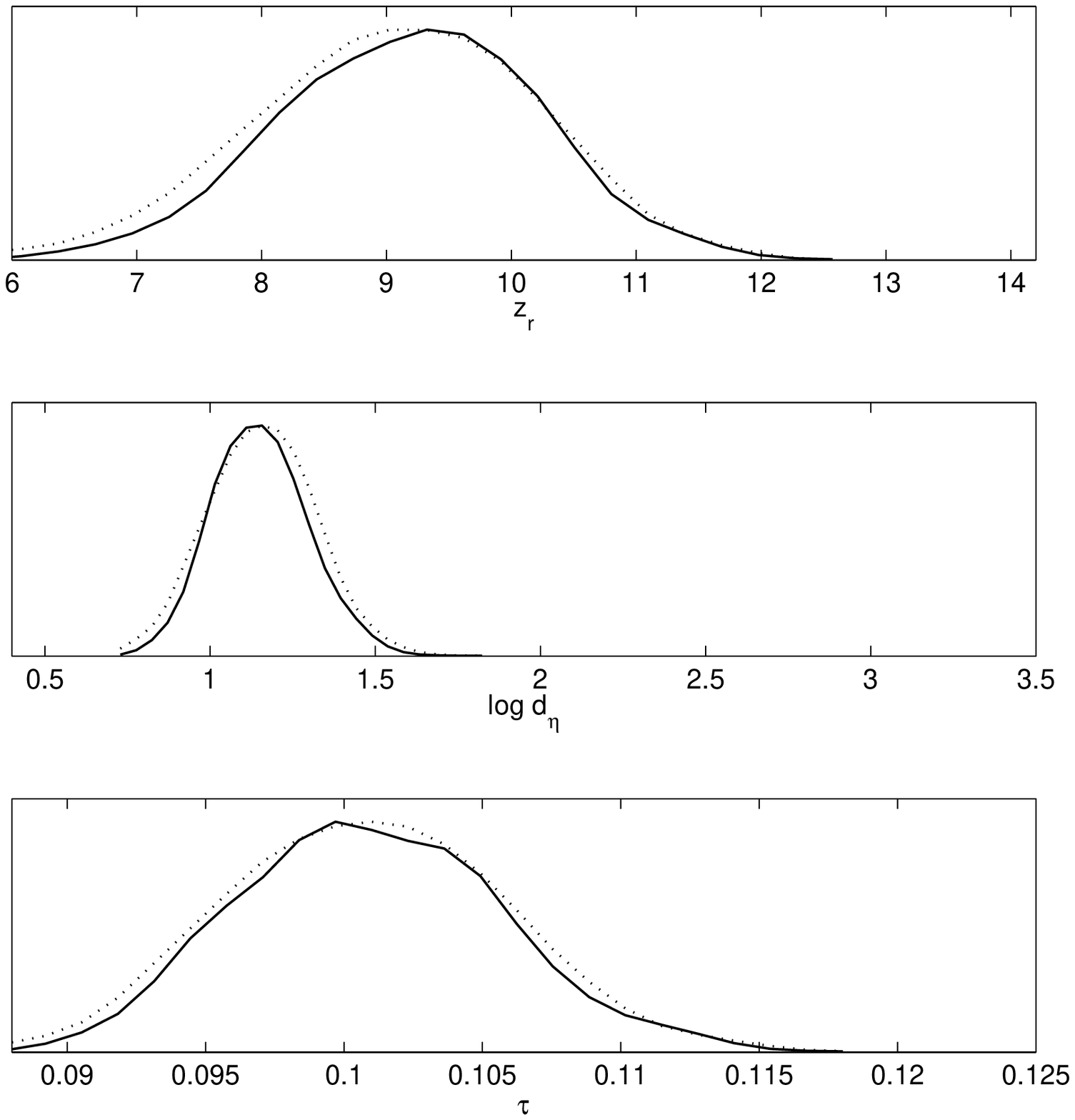}  
\end{tabular}
\caption{\label{f:paramscvl} As Figure~\ref{f:paramspl}, for a cosmic
variance limited experiment. The axes scales are the same as in 
Figure~\ref{f:paramspl}.}
\end{figure*}

The first entry in Table \ref{table1} shows that the result of
analyzing Planck data based on this chosen fiducial model with a
one-parameter instantaneous reionization model (ie. with $d_{\eta}$
held fixed at 50, varying $z_{\rm r}$ over the 6--30 prior range).
The second and third entries show the data analyzed with the (correct)
$z_{\rm r}$--$d_\eta$ model. Linear
and log priors on $d_{\eta}$ are assumed, to check for the dependence
of results on such assumptions. 

Our main result is the relative evidences of these models, where the
instantaneous reionization model has a ln evidence which is less than
2 smaller than the two-parameter models. Accordingly, Planck will not
be good enough to exclude the instantaneous reionization model, even
though the true model appears to have a quite different ionization
history. Put another way, Planck is not powerful enough to explore
two-parameter models of reionization (at least unless the true model
is even further from instantaneous reionization than our fiducial
model).

Besides the evidence, one can also compute the Bayesian complexity of
Planck data for the chosen fiducial model in the manner of Kunz,
Trotta \& Parkinson (2006). Using such an analysis Kunz et al. found
that $\tau$ is already a required parameter with WMAP 3-yr data (and a
similar analysis would likely show that another reionization parameter
is required when the evidence supports it).

This is, however, not quite the end of the story. The parameter
distributions given from the nested sampling algorithm are shown in
Figure~\ref{f:paramspl}.  It is apparent from this that the estimated
$\tau$ (and $z_{\rm r}$) are biased high in the instantaneous
reionization model.  The bias goes away in the two-parameter model, as
it should since that model can describe the true behaviour of the
data. Accordingly, to avoid a possible bias in measuring $\tau$ one
should consider both the one-parameter and two-parameter models, and
Bayesian model average as in Liddle et al.~(2006b) to obtain
constraints on $\tau$ (the cost being a slightly increased uncertainty
in $\tau$). Similar conclusions have been reported in other papers
(eg. Kaplinghat et al. 2003; Holder et al. 2003).

\begin{table*}
\centering
\begin{tabular}{|c|c|c|c|}
\hline
Model, priors & parameter estimates & ln Evidence & $\Delta$ ln E \\ 
\hline \hline 
(incorrect) $z_{\rm max}=20$ & $z_{\rm r}=8.8\pm1.4$,
$d_\eta=1.7\pm0.6$ & 
$-8.0\pm0.1$ & 7.8 \\
$z_{\rm r}$: 6--20, $d_\eta$: 0--10 &  $\tau=0.100\pm0.004$ & & \\
(incorrect) $z_{\rm max}$--$x_e$ model & $z_{\rm max}=18.9\pm0.8$,
$x_e=0.39\pm0.05$ & $-10.8\pm0.1$ & 5.0 \\
$z_{\rm max}$: 7--30, $x_e$: 0--1 & $\tau=0.094\pm0.003$ & & \\ \hline
\end{tabular}
\caption{As Table~\ref{table1}, but
analyzing the same fiducial model with an
incorrect test model, for the cosmic variance limited case. The
$\Delta \ln E$ are with respect to the instantaneous model in
Table~\ref{table1}.} 
\label{table2}
\end{table*}

We have made some assumptions about the true (fiducial) model that we
are not certain about. In practice we don't know the fiducial $z_{\rm
max}$ when reionization started. This has been assumed to be 30 in the
fiducial model. A different $z_{\rm max}$ corresponds to a different
reionization history, hence $z_{\rm max}$ could be treated as an
additional reionization parameter, but we don't go into a third
reionization parameter here. Instead we ask what outcome arises if we
analyze data so simulated (with a $z_{\rm max}$ of 30) with a $z_{\rm
r}$--$d_\eta$ model with $z_{\rm max}=20$. Such an `incorrect' model
would not be distinguishable from the true model by Planck. Further,
if our incorrect model was not a smooth transition model but one
involving a step function, again with only two parameters,
corresponding to $z_{\rm max}$ (prior range 7--30), with reionization
ending at redshift 6, and with a constant reionization fraction in
between these two redshifts of $x_e$ (prior range 0--1), such a model
would again not be distinguishable from the assumed true model by
Planck.  These results are borne out of numbers presented in the next
subsection for a cosmic variance limited hypothetical experiment.

\subsection{Cosmic variance limited case} 

For a cosmic variance limited hypothetical experiment, the
corresponding results are shown in the lower panel of
Table~\ref{table1} and in Figure~\ref{f:paramscvl}. Again only TE and
EE spectra are considered out to a maximum multipole of 100.  This
time the evidence favours the smooth and gradual transition $z_{\rm
r}$--$d_\eta$ model decisively over the instantaneous reionization
model.

These results also show that, as before, a simpler model leads to a
biased $\tau$, a bias that disappears upon using a complicated enough
model for reionization. The choice of prior on $d_\eta$ (log or
linear) doesn't make much difference.

Table~\ref{table2} shows an incorrect assumption regarding
$z_{\rm max}$ of the $z_{\rm r}$--$d_\eta$ model does not
significantly affect the evidence, ie. $z_{\rm max}=20$ is
indistinguishable from $z_{\rm max}=30$.  $d_{\eta}$ is underestimated
to make up for the difference in $z_{\rm max}$, and $\tau$ is not
misestimated. It also shows that the smooth and gradual transition
model is favoured strongly, almost decisively, over the incorrect step
model based on the difference in log evidence, while $\tau$ is biased
low under this incorrect model assumption.  Both incorrect models are
clearly distinguishable from (and favoured over) the instantaneous
reionization model. However the fact that different choices of $z_{\rm
  max}$ are not distinguishable indicates that even cosmic variance
limited experiments cannot probe very fine details of the reionization
history. 

\section{Conclusions}

We find that Planck is not expected to be able to distinguish
significantly between a single-parameter and a two-parameter model of
reionization in the model comparison sense, though it will mildly
favour the two-parameter model for our chosen fiducial values. If the
parameter values of the two-parameter true model were more extreme,
then Planck might favour it significantly. However Bayesian model
averaging the parameters over the two models will eliminate any bias
in the optical depth to rescattering.

A cosmic variance limited hypothetical experiment will be able to
decisively distinguish between the one- and two-parameter models, to
distinguish between some two-parameter models, and may be able to
go onto a third parameter.

The model comparison approach advocated here should if possible be
applied to models parameterized by describing physically relevant
quantities for the reionization history model; the phenomenological
quantities employed here are an intermediate step

\section*{Acknowledgments}

We thank Richard Battye, Jochen Weller, and Yun Wang for discussions.

\bsp

\end{document}